\documentclass [11pt,a4paper] {article}

\usepackage[cp1252]{inputenc}
\usepackage[english]{babel}

\usepackage{amssymb}
\usepackage{amsmath}
\usepackage{amsfonts,amssymb}

\usepackage[dvips]{graphicx}

\DeclareMathAlphabet{\mathpzc}{OT1}{pzc}{m}{it}

\begin{document}

\title{Effectively calculable quantum mechanics}

\author{Arkady Bolotin\footnote{$Email: arkadyv@bgu.ac.il$} \\ \textit{Ben-Gurion University of the Negev, Beersheba (Israel)}}

\maketitle

\begin{abstract}\noindent According to mathematical constructivism, a mathematical object can exist only if there is a way to compute (or ``construct'') it; so, what is non-computable is non-constructive. In the example of the quantum model, whose Fock states are associated with Fibonacci numbers, this paper shows that the mathematical formalism of quantum mechanics is non-constructive since it permits an undecidable (or effectively impossible) subset of Hilbert space. On the other hand, as it is argued in the paper, if one believes that testability of predictions is the most fundamental property of any physical theory, one need to accept that quantum mechanics must be an effectively calculable (and thus mathematically constructive) theory. With that, a way to reformulate quantum mechanics constructively, while keeping its mathematical foundation unchanged, leads to hypercomputation. In contrast, the proposed in the paper superselection rule, which acts by effectively forbidding a coherent superposition of quantum states corresponding to potential and actual infinity, can introduce computable constructivism in a quantum mechanical theory with no need for hypercomputation.\\

\end{abstract}

\section{Introduction}

\noindent Should a mathematical structure of a physical theory be \textit{algorithmic}? That is, must the collection of the mathematical objects associated with a physical theory allow the collection of all physical quantities of a particular system to determine all possible system's outcomes (or their probabilities) not only well-definably but also in \textit{an effectively calculable way}, specifically, in a finite amount of time (or in a finite number of steps provided that each step takes only a finite amount of time to perform)?\\

\noindent These questions are relevant to the ongoing foundational debate \cite{Myrvold, Lloyd2012, Wolfram, Lloyd2013}, whose main topic might be roughly expressed as follows: Are the mathematical foundations of our current physical theories necessarily non-constructive? Alternatively, are the laws of physics computable?\\

\noindent On the one hand, the requirement of effective calculability may seem to be groundless and superfluous, having no part of explaining physics. Probably because of that, current physical theories are formulated using classical analysis (including such branches as differential equations, measure theory and numerical analysis), which does not contain the requirement of effective calculability. To justify this state of affairs, one may put forward that in order to generate a result the universe does not need to proceed step by step, with a specific rule to cover what to do at each step, or to use any effective method for that matter; therefore, to claim effective calculability as a necessary property of any physical theory (that is, to allege computability of the physical laws) is to confuse the objective reality with a human way of perception, calculation or simulation of that reality.\\

\noindent On the other hand, all results produced by physical theories must be verifiable or falsifiable. Hence, if a particular physical theory gives an infinite answer to a question that should have a finite answer (of whose existence classical analysis assures us) or takes an infinite time to reach that finite answer, then this theory has a problem regarding the testability of its results, which might be a sign for a missing piece in the theory. Given that, effective calculability could be the very principle that one needs to add to the theory in question to make it testable for all possible results.\\

\noindent Of course, another argument is possible here that, for example, an ``accelerated Turing machine'' \cite{Stannett} – a model of computation that has capabilities beyond those of the standard Turing machine – could eliminate the infinite waiting time from the theory without requiring effective calculability but at the cost of admitting infinitely short times for performing each step during the computation. But then again, given a widely believed breakdown of space-time structure below the level of the Planck time, allowing such infinitely short times in the theory may appear to be as much unphysical as the infinite waiting time itself.\\

\noindent Such a course of reasoning makes evident that assuming or rejecting effective calculability has to be considered just as any other assumption or axiom of the mathematical framework of a physical theory and consequently treated as such. In other words, in any attempt to examine the question whether or not it is true that the laws of physics are computable, one must elucidate all the conclusions or consequences that would be brought about in the physical theory by the postulation or rejection of effective calculability.\\

\noindent The main goal of this paper is to do exactly this, viz., \textit{to demonstrate the consequences of the acceptance of effective calculability in quantum mechanics}. Henceforth in the paper by ``quantum mechanics'' we will refer collectively to all theories accounting for quantum phenomena, such as the ``standard quantum mechanics'' introduced by W. Heisenberg and E. Schr\"odinger in 1925–1926, in opposition, for example, to ``Collapse Theories'' or ``Bohmian mechanics'' that are mathematically different theories, rather than different interpretations of quantum mechanics \cite{Styer}.\\

\section{A quantum model whose number states are associated with Fibonacci numbers}\label{Model}

\noindent Let us start by considering the following linear equation with unknown numbers $x_1$, $x_2$, and $x_3$:\\

\begin{equation} \label{Diophantine_equation}
      D\!\left(x_1,x_2,x_3\right)=x_3-x_2-x_1=0
      \;\;\;\;  .
\end{equation}
\smallskip

\noindent To find out whether this equation has a non-negative integer solution by \textit{quantum algorithms}, it requires the realization of a Fock space \cite{Altland} – i.e., the sum of a set of Hilbert spaces representing number states with well-defined numbers of particles. On this Fock space, we construct the quantum Hamiltonian $H\!_{D}$ corresponding to the equation $D\!\left(x_1,x_2,x_3\right)$:\\

\begin{equation} \label{Hamiltonian}
      H\!_{D}=\left(
                                  a^{\dagger}_{3}a_{3}
                                - a^{\dagger}_{2}a_{2}
                                - a^{\dagger}_{1}a_{1}
                     \right)^2
      \;\;\;\;  .
\end{equation}
\smallskip

\noindent where the creation $a^{\dagger}_j$ and annihilation $a_j$ operators similar to those of the three-dimensional quantum harmonic oscillator\\

\begin{equation} \label{Oscillator}
     j,k \in \{1,2,3\}\!: \quad
     \begin{array}{l l}
         \lbrack{a_j, a^{\dagger}_{k}}\rbrack
         \equiv
         a_j a^{\dagger}_{k} - a^{\dagger}_{k}a_j = \delta_{jk} \;\;\;\;   , \\
         
         \lbrack{a^{\dagger}_j, a^{\dagger}_{k}}\rbrack
          =
         \lbrack{a_j, a_{k}}\rbrack = 0  \;\;\;\;   ,
      \end{array}
      \;\;\;\;   
\end{equation}
\smallskip

\noindent make up the number operators $N_j$\\

\begin{equation} \label{Number_operator} 
     \begin{array}{l l}
         N_j \equiv a^{\dagger}_{j}a_j \;\;\;\;   , \\
         \lbrack N_j,H\!_{D}\rbrack =\lbrack N_j,N_k\rbrack=0 \;\;\;\;   ,
      \end{array}
      \;\;\;\;   
\end{equation}
\smallskip

\noindent which have only non-negative integer eigenvalues $n_j$ and whose eigenstates $\left|\!\left.{\psi} \!\right.\right\rangle$ are those of the Hamiltonian $H\!_{D}$\\

\begin{equation} \label{Eigenstates} 
     \begin{array}{l l}
         N_j\left|\!\left.{\psi} \!\right.\right\rangle = n_j \! \left|\!\left.{\psi} \!\right.\right\rangle                                                 \;\;\;\;   , \\
         H\!_{D}\left|\!\left.{\psi} \!\right.\right\rangle = \left(n_1-n_2-n_3\right)^2 \left|\!\left.{\psi} \!\right.\right\rangle \;\;\;\;   .
      \end{array}
      \;\;\;\;   
\end{equation}
\smallskip

\noindent In this way, performing a projective measurement of the ground energy $E\!_{D}$ of the quantum system governed by the Hamiltonian (\ref{Hamiltonian}), one can answer whether or not the Diophantine equation (\ref{Diophantine_equation}) has an integer solution $n_3-n_2-n_1=0$.\\

\noindent In principle the equation (\ref{Diophantine_equation}) may have infinitely many integer solutions, so the zero ground state $\left|\!\left.{\psi_{0}} \!\right.\right\rangle$ of the Hamiltonian (\ref{Hamiltonian}) (i.e., the state with the zero ground energy $E\!_{D}=0$) will be a linear superposition of Fock states (that is, a superposition of states with definite particle number)\\

\begin{equation} \label{Superposition} 
       \left|\!\left.{\psi_{0}} \!\right.\right\rangle
        =
        \sum_{i=1}^{\infty} c_i 
                                                \left|\!\left.{n_{1_i}} \!\right.\right\rangle\!
                                                \left|\!\left.{n_{2_i}} \!\right.\right\rangle\!
                                                \left|\!\left.{n_{3_i}} \!\right.\right\rangle
      \;\;\;\;   
\end{equation}
\smallskip

\noindent where $n_{j_i}$ specifies the number of particles in the $i$-th state $j_i$, while the superposition coefficients $c_i$ meet the normalization requirement $\sum_{i=1}^{\infty}|c_i|^2=1$. Among the non-vacuum states $\left|\!\left.{n_{1_i}} \!\right.\right\rangle\!\left|\!\left.{n_{2_i}} \!\right.\right\rangle\!\left|\!\left.{n_{3_i}} \!\right.\right\rangle$ (with nonzero number of particles) one may find such that\\

\begin{equation} \label{Fibonacci_numbers} 
     \begin{array}{l l}
         n_{1_i}=F_{1_i} \;\;\;\;   , \\
         n_{2_i}=F_{2_i} \;\;\;\;   , \\
         n_{3_i}=F_{3_i} \;\;\;\;   , 
      \end{array}
      \;\;\;\;   
\end{equation}
\smallskip

\noindent where $F_{1_i}$, $F_{2_i}$, and $F_{3_i}$ are sequential Fibonacci numbers\\

\begin{equation} \label{Fibonacci_relation}
      F_{3_i}=F_{1_i}+F_{2_i}
      \;\;\;\;   
\end{equation}
\smallskip

\noindent (in the vacuum state $\left|\!\left.{0_{1_i}} \!\right.\right\rangle\!\left|\!\left.{0_{2_i}} \!\right.\right\rangle\!\left|\!\left.{0_{3_i}} \!\right.\right\rangle$ all $F_{j_i}$ are the same and equal to zero); let us denote such states as \textit{Fibonacci states} $\left|\!\left.{F_{1_i}} \!\right.\right\rangle\!\left|\!\left.{F_{2_i}} \!\right.\right\rangle\!\left|\!\left.{F_{3_i}} \!\right.\right\rangle$.\\

\noindent Since the set of natural numbers $\mathbb{N}$ can be written as the direct sum $\mathbb{N}=F \oplus Z$ of two of its proper subsets, the Fibonacci $F$ and non-Fibonacci $Z$ numbers, the eigenspace $\mathcal{E}_0$ of the zero ground energy $E\!_{D}=0$ for the considered quantum model can be expressible as the direct sum of two subsets $\mathcal{E}_F$ and $\mathcal{E}_Z$ formed by the Fibonacci and non-Fibonacci states, respectively,\\

\begin{equation} \label{Eigenspace} 
       \mathcal{E}_0 = \mathcal{E}_F \oplus \mathcal{E}_Z
        =
       \{ \left|\!\left.{F_{1_i}} \!\right.\right\rangle\!\left|\!\left.{F_{2_i}} \!\right.\right\rangle\!\left|\!\left.{F_{3_i}} \!\right.\right\rangle \}
       \oplus
       \{ \left|\!\left.{z_{1_i}} \!\right.\right\rangle\!\left|\!\left.{z_{2_i}} \!\right.\right\rangle\!\left|\!\left.{z_{3_i}} \!\right.\right\rangle \}
      \;\;\;\;  ,
\end{equation}
\smallskip

\noindent where the non-Fibonacci states are defined by\\

\begin{equation} \label{non_Fibonacci} 
       \left|\!\left.{z_{1_i}} \!\right.\right\rangle\!\left|\!\left.{z_{2_i}} \!\right.\right\rangle\!\left|\!\left.{z_{3_i}} \!\right.\right\rangle
        \in
       \{ \left|\!\left.{n_{1_i}} \!\right.\right\rangle\!\left|\!\left.{n_{2_i}} \!\right.\right\rangle\!\left|\!\left.{n_{3_i}} \!\right.\right\rangle \}
       \backslash
       \{ \left|\!\left.{F_{1_i}} \!\right.\right\rangle\!\left|\!\left.{F_{2_i}} \!\right.\right\rangle\!\left|\!\left.{F_{3_i}} \!\right.\right\rangle \}
      \;\;\;\;   
\end{equation}
\smallskip

\noindent and the vacuum state $\left|\!\left.{0_{1_i}} \!\right.\right\rangle\!\left|\!\left.{0_{2_i}} \!\right.\right\rangle\!\left|\!\left.{0_{3_i}} \!\right.\right\rangle$ belongs to the intersection $\mathcal{E}_F \cap \mathcal{E}_Z$; subsequently, the system's zero ground state $\left|\!\left.{\psi_{0}} \!\right.\right\rangle$ can be presented as the superposition of the Fibonacci and non-Fibonacci states\\

\begin{equation} \label{Superposition2} 
       \left|\!\left.{\psi_{0}} \!\right.\right\rangle
        =
        c_i \left|\!\left.{0_{1_i}} \!\right.\right\rangle\!\left|\!\left.{0_{2_i}} \!\right.\right\rangle\!\left|\!\left.{0_{3_i}} \!\right.\right\rangle
        +
        \sum_{k} {\alpha}_k  \left|\!\left.{F_{1_k}} \!\right.\right\rangle\!\left|\!\left.{F_{2_k}} \!\right.\right\rangle\!\left|\!\left.{F_{3_k}} \!\right.\right\rangle
        +
        \sum_{l} {\beta}_l  \left|\!\left.{z_{1_l}} \!\right.\right\rangle\!\left|\!\left.{z_{2_l}} \!\right.\right\rangle\!\left|\!\left.{z_{3_l}} \!\right.\right\rangle
      \;\;\;\;   
\end{equation}
\smallskip

\noindent such that $c_i$ and the coefficients ${\alpha}_k$ and ${\beta}_l$ before the non-vacuum states satisfy the condition $c_i,{\alpha}_k,{\beta}_l \in \{c_m\}_{m=1}^{\infty}$.\\

\noindent It is natural to ask whether the Fibonacci states subset $\mathcal{E}_F$ is recognizable. Explicitly, given a positive triple $(b_1,b_2,b_3)$ gotten through the measurement on the zero ground state of the Hamiltonian (\ref{Hamiltonian}), can one decide in a finite amount of time whether or not its elements $b_1,b_2,b_3$ are Fibonacci numbers?\\

\section{Recognizing Fibonacci numbers}\label{Recognizing}

\noindent A straightforward (brute-force) way to recognize Fibonacci numbers is to generate them until one becomes equal to a given positive integer $b_j$: If it does, then the integer $b_j$ is a Fibonacci number, if not, the numbers will eventually become bigger than $b_j$, and the procedure will stop.\\

\noindent Another way is to use the closed-form expression for Fibonacci numbers known as Binet's formula \cite{Livio, Seroul}. According to this expression, the positive integer $b_j$ would belong to the Fibonacci sequence if and only if the closed interval $S_j$ defined by\\

\begin{equation} \label{Interval} 
       S_j
        =
        \left [ 
           {\phi}b_j  - \frac{1}{b_j}
           ,
           {\phi}b_j  + \frac{1}{b_j}
        \right ] 
      \;\;\;\;  ,
\end{equation}
\smallskip

\noindent where $\phi$ is the golden ratio\\

\begin{equation} \label{Golden_ratio} 
       \phi
        =
        \frac{1}{2} \left( 1 + \sqrt{5} \right) 
      \;\;\;\;  ,
\end{equation}
\smallskip

\noindent intersects the set of all natural numbers $\mathbb{N}$ at some element (or elements), that is,\\

\begin{equation} \label{Criterion} 
       S_j \cap \mathbb{N} \neq \emptyset
      \;\;\;\;  .
\end{equation}
\smallskip

\noindent Let the golden ratio $\phi=1+\{\phi\}$, where $\{\phi\}$ denotes the infinite continued fraction\\

\begin{equation} \label{Continued_fraction} 
       \{\phi\} = \cfrac{1}{1
                                + \cfrac{1}{1
                                   + \cfrac{1}{1 
                                     + \cdots 
                                      } 
                                   }
                                }
                                =[{0;1,1,1,\dots}] 
      \;\;\;\;  ,
\end{equation}
\smallskip

\noindent be calculated to the accuracy of the $n^\mathrm{th}$  Diophantine approximation of $\{\phi\}$\\

\begin{equation} \label{Approximation} 
      \{\phi\}
                        \cong 
                                     \big{\lbrack} {0;\underbrace{1,1,1,\dots,1}_n} \,\big{\rbrack} = \frac{p_n}{q_n}
      \;\;\;\;  ,
\end{equation}
\smallskip

\noindent such that the positive integers $p_n$ and $q_n$ are given by the Fibonacci recurrence relation\\

\begin{equation} \label{Recurrence_relation} 
     \begin{array}{l l}
         p_n = q_{n-1} \;\;\;\;   , \\
         q_n =q_{n-1} + q_{n-2} \;\;\;\;    
      \end{array}
      \;\;\;\;   
\end{equation}
\smallskip

\noindent with the seed values $p_1 =1$ and $q_1 =1$ (as it can be seen, the denominator $q_n$ increases strictly monotonic when $n$ goes up, i.e., when additional unities are included in the approximation of $\{\phi\}$; just observe, for example, the first four approximations of the fraction $\{\phi\}$: $[0;1]=1/1$, $[0;1,1]=1/2$, $[0;1,1,1]=2/3$, $[0;1,1,1,1]=3/5$). Then, the criterion (\ref{Criterion}) can be rewritten in the form of the following equality \cite{Bolotin}:\\

\begin{equation} \label{Equality} 
       \bigg\lfloor
            b_j \frac{p_n}{q_n} + b_j + \frac{1}{b_j}
        \bigg\rfloor
        -
       \bigg\lceil
            b_j \frac{p_n}{q_n} + b_j - \frac{1}{b_j}
        \bigg\rceil
         =
        0
      \;\;\;\;  ,
\end{equation}
\smallskip

\noindent where $\lfloor\cdot\rfloor$ and $\lceil\cdot\rceil$ stand for the floor and ceiling functions, respectively.\\

\noindent Suppose that for the positive triple $(b_1,b_2,b_3)$ measured on the zero ground state $\left|\!\left.{\psi_{0}} \!\right.\right\rangle$ of the Hamiltonian (\ref{Hamiltonian}) the equality (\ref{Equality}) does hold. To decide whether in this case $b_1,b_2,b_3$ are indeed Fibonacci numbers (and correspondingly the system's quantum state after the measurement is a Fibonacci state), the upper bound for the Diophantine approximations ${p_n}/{q_n}$  of $\{\phi\}$ \cite{Hailperin, Waldschmidt}\\

\begin{equation} \label{Upper_bound} 
      \bigg|
      \{\phi\} - \frac{p_n}{q_n} 
      \bigg|
      <
      \frac{1}{\sqrt{5}{q_n}^2}
      \;\;\;\;   
\end{equation}
\smallskip

\noindent must be much less than the reciprocals of the integers $b_1,b_2,b_3$, meaning that the fraction $\{\phi\}$ must be calculated to such an accuracy that the following inequality holds\\

\begin{equation} \label{Inequality} 
      {q_n}^2
      \gg
      \frac{b_j}{\sqrt{5}} 
      \;\;\;\;  .
\end{equation}
\smallskip

\noindent With regard to the last inequality, it is important to note two things.\\

\noindent First, in contrast to any other irrational number $\gamma$, for which there are infinitely many Diophantine approximations $p_n/q_n$  whose distance from $\gamma$ is significantly smaller than the limit ${1}/{\sqrt{5}{q_n}^2}$, for the golden ratio fraction $\{\phi\}$ the upper bound ${1}/{\sqrt{5}{q_n}^2}$ is tight: Any Diophantine approximation of $\{\phi\}$ almost exactly keeps this distance away from $\{\phi\}$ (which makes the golden ratio $\phi$ the most difficult number to approximate rationally) \cite{Hardy}.\\

\noindent Second, since the zero ground state $\left|\!\left.{\psi_{0}} \!\right.\right\rangle$ of the Hamiltonian (\ref{Hamiltonian}) is formed by the superposition of all possible Fibonacci and non-Fibonacci states, measuring the triple $(b_1,b_2,b_3)$ can yield any of the results $b_{1_i},b_{2_i},b_{3_i} \in \mathbb{N}$ with corresponding probabilities given by ${|c_i |}^2$. Thus, in the most general case, $b_j$ might be anywhere from zero to infinity.\\

\noindent Together these two things indicate that in order to recognize correctly number states of the Fibonacci subset $\mathcal{E}_F$  included in the eigenspace $\mathcal{E}_0$  of the considered quantum system (i.e., to decide correctly whether those states are Fibonacci or not) is necessary to calculate the fraction $\{\phi\}$ to an unbounded accuracy ${p_{\infty}}/{q_{\infty}}$, which can certainly be achieved only by way of applying the recurrence relation (\ref{Recurrence_relation}) infinitely many times and hence would take an infinite amount of time (using the brute-force method described at the beginning of this Section would involve generating the entire Fibonacci sequence, which would obviously take an infinite time too).\\

\noindent Such an infinite waiting time, however, presents a problem to the mathematical formalism of quantum mechanics: Namely, when a complete description of a quantum state is given in the form of a Fibonacci state or arbitrary superpositions of Fibonacci states, it is principally impossible to always verify this – i.e., to decide in every measurement whether or not the given state is Fibonacci – since it might demand an infinite amount of time. But this constitutes a contradiction to the prevailing conception of any physical theory that must express \textit{only those predictions, which can be testable in all cases} (albeit even in principle). So, how does it come to be that quantum mechanics predicts something that cannot be verified even in theory?\\

\section{Ways to resolve the problem}\label{Ways}

\noindent Let us see how this problem can be resolved.\\

\subsection{Fibonacci numbers have no physical relevance}

\noindent To begin with, one can merely object to the existence of any problem here asserting that the Fibonacci sequence is a mathematical object, which does not correspond to any actual process or a physical system, and, as a result, recognizing the Fibonacci numbers does not have a lot more meaning in the physical world than, say, recognizing the odd numbers. Therefore, the considered above quantum model whose states are associated with Fibonacci numbers is just a ``toy model'' that has nothing to do with the physical realm.\\

\noindent Still, even if one dismisses that the Fibonacci numbers appear in nature often enough to prove that they reflect some naturally occurring patterns (particularly, phyllotaxic patterns generated whenever a vascular plant repeatedly produces similar botanical elements at its tip such as leaves, bractae, florets etc.; these patterns are directly related to the Fibonacci sequence and the golden ratio and in fact are so regular that a physicist can compare their order to that of crystals; see for example paper \cite{Douady} that investigates the striking predominance of Fibonacci order in botany), the problem won't go away completely.\\

\noindent The problem created by unrecognizability of the Fibonacci states subset $\mathcal{E}_F$ in a finite time might still be important to the application of quantum formalism to so-called \textit{quantum-like systems}, i.e., non-physical systems ranging, for example, from finance \cite{Baaquie, Haven2005} and population dynamics \cite{Bagarello} to social science \cite{Haven2013}, psychology \cite{Aerts}, cognition \cite{Busemeyer} and neuroscience \cite{Khrennikov}.\\

\subsection{Physically realizable integers are limited in size}

\noindent Seeing as the assumption of infinite quantities is apparently never realized in the observable universe, one can conclude that all the integers that are related to natural processes are limited in size. Conforming to such a finistic conclusion (which is in line with the mathematical philosophy of finitism \cite{Ye} and especially the theory of explicit finitism \cite{Kornai}), for a physically meaningful quantum system the results of the measurement of the triple $(b_1,b_2,b_3)$ on the zero ground state $\left|\!\left.{\psi_{0}} \!\right.\right\rangle$ of the Hamiltonian (\ref{Hamiltonian}) would always be in a finite interval and, hence, recognizable as the Fibonacci or non-Fibonacci numbers in a finite amount of time.\\

\noindent Let us consider the computable function $f$, which equals 1 if $b_j$ belongs to the Fibonacci sequence and zero otherwise:\\

\begin{equation} \label{Computable_function} 
      f\!\left(b_j\right) = \bigg\lbrace{
                    \begin{array}{l l}
                                      1, & \quad b_j \in \{F_m\}^{\infty}_{m=1}\\
                                      0, & \quad \mbox{else}\\
                     \end{array}}
      \;\;\;\;  .
\end{equation}
\smallskip

\noindent As it can be readily seen, if the physically realizable positive integer $b_j$ were to be limited in size, then there would exist a naturally originated limit on computability of the function $f\!\left(b_j\right)$.\\

\noindent Unfortunately, it is very hard for the finistic conclusion to answer the charge of \textit{arbitrariness}: No matter where the limit on computability would be drawn (say, as it is proposed in the paper \cite{Kornai}, it would be put at the level of the Ackermann function \cite{Monin} $A(4,4)=2^{2^{2^{65536}}}-3$), it would be always \textit{ad hoc} and so perpetually subject to shifting. Accordingly, one cannot modify the function $f\!\left(b_j\right)$ so that to accommodate this limit and at the same time preserve the procedure for computing the function $f\!\left(b_j\right)$ well-defined. This means that the given finistic conclusion is logically deficient as there is no way to formulate it unambiguously.\\

\subsection{Hypercomputation}

\noindent Assume that the mathematical formalism of quantum mechanics is \textit{complete} (i.e., no additional hypothesis need to be admitted to its foundation) and applicable to any physical system. Then, as it follows from the Section \ref{Recognizing}, to guarantee testability of all predictions made within the frame of the quantum formalism, the function $f\!\left(b_j\right)$ must be computable for any unlimited arguments $b_j$ in a finite amount of time. That might be only if this function $f\!\left(b_j\right)$ were to be computable either non-recursively or by ``super-Turing'' machines.\\

\noindent To be sure, if it were possible to find the exact value of the fraction $\{\phi\}$ either without applying the recurrence relation (\ref{Recurrence_relation}) infinitely many times (say, through the use of a computing device, such as a BSS machine \cite{Schonhage}, which has the ability to compute $x+y$, $x-y$, $xy$, $x/y$, and  $\lfloor x\rfloor$ in a single step for any two infinite-precision real numbers $x$ and $y\ne0$) or with calculating this relation on every occasion of $n$ in an unboundedly short time-length (say, by using an infinite time Turing machine that includes as a part the accelerating Turing machine mentioned in the Introduction), then the function $f\!\left(b_j\right)$ could be definitely computable for any $b_j$ in a finite amount of time.\\

\noindent Yet, real computers (operating on the set of real numbers), infinite time Turing machines, or all other models of hypercomputation proposed so far do not seem to be physically constructible and reliable (at least, for the moment) \cite{Davis}. This casts doubt upon the physical existence of hypercomputers and, in this way, upon the assumption of completeness of the quantum formalism (which brings into being the need for hypercomputation).\\

\subsection{Effectively calculable quantum mechanics}

\noindent So, as an alternative, let us assume that the mathematical formalism of quantum mechanics is not complete in such a way that the requirement of effective calculability has to be added to its axiomatic base in order to complete the formalism.\\

\noindent A familiar tactic to do so would be through the agency of superselection rule \cite{Giulini}.\\

\noindent Let us present the Fock space of the system we are considering – i.e., the closed set of the number states – as the direct sum of the following two superselection sectors:\\

\begin{equation} \label{Sectors} 
      \{ \left|\!\left.{n_{1_i}} \!\right.\right\rangle\!\left|\!\left.{n_{2_i}} \!\right.\right\rangle\!\left|\!\left.{n_{3_i}} \!\right.\right\rangle \}
      =
      \mathcal{H}_c \oplus \mathcal{H}_{\infty}
      \;\;\;\;  ,
\end{equation}
\smallskip

\noindent where $\mathcal{H}_c$ denotes the open set, whose each member (i.e., a quantum state) is an eigenstate of the particle number operator corresponding to a finite number of particles in the given state and can be achieved (for example, by repeatedly operating with the creation operator $a^{\dagger}_j$ on the vacuum state $a^{\dagger}_j\left|\!\left.{0_{j}} \!\right.\right\rangle=\left|\!\left.{1_{j}} \!\right.\right\rangle$, $a^{\dagger}_j\left|\!\left.{1_{j}} \!\right.\right\rangle=\left|\!\left.{2_{j}} \!\right.\right\rangle$, ...) in a finite number of steps, while $\mathcal{H}_{\infty}$ stands for the ``boundary'' set of the infinite members, i.e., the number states corresponding to an actual infinity of particles.\\

\noindent We will put forward that for all physically realizable observables $Q$ there is a superselection rule\\

\begin{equation} \label{Rule} 
      \langle {\Psi_{1}}|\,
      Q
      \left|\left.{\Psi_{2}}\!\right.\right\rangle
      =
      0
      \;\;\;\;  ,
\end{equation}
\smallskip

\noindent in the presence of which a vector of Hilbert space $\left|\!\left.{\Psi}\!\right.\right\rangle$ consisting of two components $\left|\!\left.{\Psi_{1}}\!\right.\right\rangle$ and $\left|\!\left.{\Psi_{2}}\!\right.\right\rangle$\\

\begin{equation} \label{Vector} 
      \left|\!\left.{\Psi}\!\right.\right\rangle
      =
      \frac{1}{\sqrt{2}}
      \left(
            \left|\left.{\Psi_{1}}\!\right.\right\rangle
            +
            \left|\left.{\Psi_{2}}\!\right.\right\rangle
      \right)
      \;\;\;\;   
\end{equation}
\smallskip

\noindent that belong to the two superselection sectors  $\mathcal{H}_c$ and $\mathcal{H}_{\infty}$, respectively, cannot represent a physical state. Then, substituting (\ref{Vector}) in (\ref{Rule}) will give\\

\begin{equation} \label{Substituting} 
      \langle {\Psi}|\,Q\left|\left.{\Psi}\!\right.\right\rangle
      =
      \frac{1}{\sqrt{2}}
      \left(
            \langle {\Psi_{1}}|\,Q\left|\left.{\Psi_{1}}\!\right.\right\rangle
            +
            \langle {\Psi_{2}}|\,Q\left|\left.{\Psi_{2}}\!\right.\right\rangle
      \right)
      =
      \mathrm{Tr}\!\left(\rho Q\right)
      \;\;\;\;  ,
\end{equation}
\smallskip

\noindent where the density matrix $\rho$ corresponding to the vector $\left|\!\left.{\Psi}\!\right.\right\rangle$ is given by the combination of the pure density matrices for the components $\left|\!\left.{\Psi_{1}}\!\right.\right\rangle \in \mathcal{H}_c$ and $\left|\!\left.{\Psi_{2}}\!\right.\right\rangle \in\mathcal{H}_{\infty}$\\

\begin{equation} \label{Density_matrix} 
      \rho
      =
      \frac{1}{\sqrt{2}}
      \left(
            \left|\left.{\Psi_{1}}\!\right.\right\rangle\! \langle {\Psi_{1}}|
            +
            \left|\left.{\Psi_{2}}\!\right.\right\rangle\! \langle {\Psi_{2}}|
      \right)
      \;\;\;\;   
\end{equation}
\smallskip

\noindent and therefore defines a mixed state rather than a pure state. This means that in the presence of the superselection rule (\ref{Rule}) a convex linear combination of the state vectors belonging to the superselection sectors $\mathcal{H}_c$ and $\mathcal{H}_{\infty}$ cannot be a pure state.\\

\noindent Because the (time independent) Hamiltonian $H\!_{D}$ for the considered system is the self-adjoint observable $E\!_{D}\!=\!(n_3-n_2-n_1)^2$, the Schr\"odinger evolution will never evolve a state vector of the system from one superselection sector to another, i.e., from $\mathcal{H}_c$ to $\mathcal{H}_{\infty}$, and will always evolve a pure state to a pure state. In consequence, the superposition of the physically realizable number states (\ref{Superposition}) that represents a pure state cannot contain the components in $\mathcal{H}_{\infty}$. Accordingly, measuring the triple $(b_1,b_2,b_3)$ on the zero ground state of the Hamiltonian $H_D$ can always yield only finite results $b_{1_i},b_{2_i},b_{3_i} <\infty$. Thus, in the presence of the superselection rule (\ref{Rule}) it would become principally possible (that is, effectively possible) to decide in every given measurement whether the obtained (finite) numbers $b_{1_i},b_{2_i},b_{3_i}$ are Fibonacci or not.\\

\noindent As it can be seen, the superselection rule (\ref{Rule}) is equivalent to the assumption that the matrix elements of the physically realizable observables $Q$ cannot distinguish between states from the superselection sectors $\mathcal{H}_c$ and $\mathcal{H}_{\infty}$, that is, between \textit{potential (computational) infinity} (such as a non-terminating process of consecutively applying the creation operator $a^{\dagger}_j$ to the vacuum state $\left|\!\left.{0_{j}} \!\right.\right\rangle$) \textit{and actual infinity} (such as the set of all natural numbers $\mathbb{N}$) \cite{Fraenkel}. In other words, the superselection rule (\ref{Rule}) postulates that \textit{in the physical universe a coherent superposition of states corresponding to potential and actual infinity cannot be verified or prepared}.\\

\noindent The fact that no one has ever succeeded in forming such a superposition can provide some evidence for the superselection rule (\ref{Rule}). Again, the apparent absence of infinite things within the region where all scientific experiments and human experiences happen can be a further indication lending support to this rule.\\

\noindent The question, nonetheless, remains about how this proposed superselection rule could be understood: Namely, is it a formalistic mathematical device or full of a physical meaning?\\

\noindent Let the volume of a system be taken to grow in proportion with the number of particles in the system; then, the actual infinity of particles would correspond to a system occupying an infinite volume of space. Assuming that such a system may exist (which is equivalent to the assumption that the universe – while continuing to expand exponentially on the largest scales – is already spatially infinite \cite{Aguirre}), the physical reason keeping a coherent superposition of states relating to the potential and actual infinity of particles from occurring might be the presence of new physics at infinitely long distances (or ones that at least large than $10^{10^{10^{122}}}$ megaparsecs \cite{Page}).\\

\section{Concluding remarks: Computable constructivism in quantum theory}\label{Concluding}

\noindent Ideologically, the effectively calculable quantum mechanics approach outlined above is closely related to \textit{mathematical constructivism}, which asserts that a mathematical object exists only if there is a way (i.e., an effective procedure) to compute (or ``construct'') it and, accordingly, what is non-computable is non-constructive \cite{Troelstra}.\\

\noindent In view of that, the mathematical formalism of quantum mechanics should be considered \textit{non-constructive}\footnote{\noindent It is noteworthy that non-constructivism of quantum mechanics (in either the sense of \textit{intuitionism} or that of \textit{Bishop-constructivism}) was already demonstrated in the paper \cite{Hellman}, which argued that unbounded linear Hermitian operators in Hilbert space are not even legitimately recognizable as mathematical objects from a thoroughgoing constructivist point of view.} since it permits a subset of Hilbert space that is effectively impossible (i.e., noncomputable): As it has been demonstrated, this formalism allows the existence of the Fibonacci states subset in the Fock space of the quantum model (whose Hamiltonian mimics the form of the left–hand–side squared of the Diophantine equation for non-negative integers) such that there is no algorithm that can in a finite amount of time decide whether or not an arbitrary state of the model belongs to this subset.\\

\noindent At the same time, if one believes that verifiability/falsifiability is the most crucial property of any physical theory, one need to accept that quantum mechanics must be an effectively calculable and so mathematically constructive theory.\\

\noindent Therewith, a way to introduce mathematical (to be exact, \textit{computable}) constructivism in quantum mechanics without revising its mathematical foundation leads to hypercomputation, that is, to the idea that physical systems can be identified or designed (constructed or exploited), which can compute non-recursive functions or outperform the standard Turing machines.\\

\noindent In contrast, the proposed superselection rule, which acts by effectively forbidding a coherent superposition of quantum states that correspond to potential and actual infinity, institutes computable constructivism in a quantum mechanical theory with no need for hypercomputation.\\

\end{document}